\documentclass[aps,prb,twocolumn,superscriptaddress]{revtex4-2}
\usepackage[hidelinks,linktoc=all]{hyperref} 
\usepackage{here}
\usepackage{soul}
\usepackage{siunitx}
\usepackage{braket}
\usepackage[dvipdfmx]{graphicx}
\usepackage[version=4]{mhchem}
\begin{document}
\title{Bayesian model comparison of type-I and type-II
ultrafast demagnetization dynamics}
\author{Hiroki Wadati}
\email{wadati@sci.u-hyogo.ac.jp}
\affiliation{Department of Material Science, Graduate School of Science, University of Hyogo, Ako, Hyogo 678-1297, Japan}
\affiliation{Institute of Laser Engineering, The University of Osaka, Suita, Osaka 565-0871, Japan}
\author{Tetsuro Ueno}
\affiliation{Synchrotron Radiation Research Center, Kansai Institute for Photon Science, National Institutes for Quantum Science and Technology, Sayo, Hyogo 679-5148, Japan}
\affiliation{Quantum Materials and Applications Research Center, Takasaki Institute for Advanced Quantum Science, National Institutes for Quantum Science and Technology, Takasaki, Gunma 370-1292, Japan}
\begin{abstract}
Ultrafast demagnetization dynamics are often phenomenologically classified into type-I and type-II responses according to their temporal evolution following femtosecond laser excitation. However, finite experimental temporal resolution and noise can substantially obscure the intrinsic dynamics and complicate this classification. In this work, we investigate the distinguishability of type-I and type-II demagnetization dynamics using Gaussian-convolved phenomenological models and Bayesian information criterion-based statistical model comparison. Synthetic datasets with varying temporal resolution and noise levels are first analyzed to evaluate the conditions under which the two classes can be reliably discriminated. We show that convolution with the instrumental response function significantly reduces the observable differences between the intrinsic responses, thereby producing broad regimes in which model discrimination becomes statistically inconclusive. The applicability of the framework is further demonstrated through analysis of representative experimental ultrafast demagnetization data from NiCo$_2$O$_4$ thin films. These results suggest that the apparent classification of ultrafast demagnetization dynamics can be highly sensitive to experimental resolution, noise level, and analysis methodology.
\end{abstract}

\maketitle

\section{Introduction}
Ultrafast demagnetization following femtosecond laser excitation has attracted sustained interest since the pioneering observation by Beaurepaire et al. \cite{1} and subsequent developments in ultrafast magnetism \cite{2}. In phenomenological descriptions of time-resolved magnetization dynamics, type-I and type-II responses are commonly discussed as distinct classes characterized by single-step and two-step demagnetization behavior, respectively \cite{3,4}. Type-I dynamics are frequently observed in elemental transition-metal ferromagnets such as Ni and Co, whereas type-II dynamics have been reported in several rare-earth, alloy, and ferrimagnetic systems \cite{3,4}. 
Although the microscopic origin of type-I and type-II dynamics remains the subject of active debate, several mechanisms have been proposed, including Elliott–Yafet spin-flip scattering [3], superdiffusive spin transport [5], and angular-momentum transfer between magnetic sublattices [6]. Depending on the material system, the relative importance of these mechanisms may vary, giving rise to different ultrafast demagnetization behaviors.

Despite its widespread adoption, the distinction between type-I and type-II demagnetization dynamics is not always straightforward in practice. Time-resolved pump-probe measurements are inevitably subject to finite temporal resolution and experimental noise, both of which can obscure subtle features of the intrinsic response, causing different phenomenological models to yield visually similar fits even when their intrinsic dynamics are qualitatively distinct. 
Conventional analyses typically rely on least-squares fitting of time-domain data using a chosen phenomenological model. However, such approaches do not provide a quantitative basis for comparing models with different structures. In regimes where the data are limited by temporal resolution or noise, this can lead to ambiguous or overconfident assignments of demagnetization type based on visual inspection or residual analysis alone. These considerations raise a fundamental question: under what conditions can time-resolved data meaningfully discriminate between type-I and type-II demagnetization dynamics? Conversely, when do experimental limitations render such classifications statistically inconclusive? 

In this work, we reformulate the classification of ultrafast demagnetization dynamics as a Bayesian model-comparison problem. Using Gaussian-convolved phenomenological models and synthetic pump-probe datasets, we investigate how temporal resolution and noise affect the distinguishability between type-I and type-II dynamics. This approach provides a quantitative criterion for identifying regimes in which model classification becomes statistically inconclusive.

The remainder of this paper is organized as follows. In Sec.~II, we introduce the phenomenological models, their convolution with a finite instrumental response function, the synthetic data generation procedure, and the Bayesian model comparison framework. Section III presents results for both synthetic benchmark datasets and representative experimental data on ultrafast demagnetization. The implications and limitations of the present approach are discussed in Sec.~IV, followed by the conclusions in Sec.~V.

\section{Model and formulation}
In this section, we define the phenomenological models for ultrafast demagnetization dynamics and their relation to experimentally observed signals. The formulation is kept general and independent of specific microscopic mechanisms, allowing it to be applied directly to a broad class of pump-probe measurements.

\subsection{Phenomenological models}
To describe ultrafast demagnetization dynamics, we consider two phenomenological models corresponding to type-I and type-II responses.

The type-I model assumes a single demagnetization channel followed by recovery dynamics:
\begin{equation}
M_{\mathrm{I}}(t)=1-A\left(1-e^{-t/\tau}\right)
e^{-t/\tau_{\mathrm{rec}}},
\label{eq:type1}
\end{equation}
where $A$ is the demagnetization amplitude, $\tau$ is the characteristic demagnetization time, and $\tau_{\mathrm{rec}}$ is the recovery time.

On the other hand, the type-II model includes both fast and slow demagnetization components:
\begin{align}
M_{\mathrm{II}}(t)
= 1 &-\Big[A_{\mathrm{fast}}
\left(1-e^{-t/\tau_{\mathrm{fast}}}\right)
\nonumber \\
&+A_{\mathrm{slow}}
\left(1-e^{-t/\tau_{\mathrm{slow}}}\right)
\Big]e^{-t/\tau_{\mathrm{rec}}}.
\end{align}
where $A_{\mathrm{fast}}$ and $A_{\mathrm{slow}}$ are the amplitudes of the fast and slow demagnetization components, respectively, and $\tau_{\mathrm{fast}}$ and $\tau_{\mathrm{slow}}$ denote the corresponding characteristic times.

To incorporate finite experimental temporal resolution, the intrinsic dynamics were convolved with a Gaussian instrumental response function (IRF),
\begin{equation}
G_{\sigma}(t)=\frac{1}{\sqrt{2\pi}\sigma}\exp\left(-\frac{t^2}{2\sigma^2}
\right),
\label{eq:gaussian_irf}
\end{equation}
where $\sigma$ represents the standard deviation of the Gaussian IRF.

The observable signal is therefore expressed as a convolution
\begin{equation}
S(t)=\int_{-\infty}^{\infty}M(t')G_{\sigma}(t-t')dt',
\label{eq:convolution}
\end{equation}
where $M(t)=$ $M_{\mathrm{I}}(t)$ or $M_{\mathrm{II}}(t)$. 

In the following analysis, the experimentally observable signals correspond to the convolved responses rather than the intrinsic dynamics themselves.

\subsection{Analytic convolution with a Gaussian IRF}
A key building block for the convolution is the response of a single exponential decay,
\begin{equation}
\Phi(t;\lambda,\sigma)=\int_{0}^{\infty}e^{-u/\lambda}\,G_\sigma(t-u)\,du,
\label{eq:phi_def}
\end{equation}
which can be evaluated analytically as
\begin{equation}
\Phi(t;\lambda,\sigma)=\frac{1}{2}\exp\!\left(\frac{\sigma^2}{2\lambda^2}-\frac{t}{\lambda}\right)
\operatorname{erfc}
\!\left(\frac{\sigma^2/\lambda - t}{\sqrt{2}\sigma}\right), 
\label{eq:phi_erfc}
\end{equation}
where $\mathrm{erfc}(x)$ denotes the complementary error function. 

Using the linearity of convolution, the observable signals corresponding to the intrinsic models in Eqs.~(1) and (2) can be written in closed form.

For the type-I model, we obtain
\begin{equation}
y_{\mathrm{I}}(t)=1-A\left[
\Phi(t; \tau_{\mathrm{rec}}, \sigma)-\Phi
\left(t; \frac{\tau \tau_{\mathrm{rec}}}
{\tau + \tau_{\mathrm{rec}}}, 
\sigma\right)\right],
\label{eq:yI}
\end{equation}
while for the type-II model,
\begin{equation}
\begin{aligned}
y_{\mathrm{II}}(t)=1 &-A_{\mathrm{fast}}
\left[\Phi(t; \tau_{\mathrm{rec}}, \sigma)
-\Phi\left(t;\frac{\tau_{\mathrm{fast}} \tau_{\mathrm{rec}}}
{\tau_{\mathrm{fast}} + \tau_{\mathrm{rec}}},
\sigma\right)\right]
\\
&-A_{\mathrm{slow}}\left[
\Phi(t; \tau_{\mathrm{rec}}, \sigma)-\Phi
\left(t;\frac{\tau_{\mathrm{slow}} \tau_{\mathrm{rec}}}
{\tau_{\mathrm{slow}} + \tau_{\mathrm{rec}}},
\sigma\right)\right].
\end{aligned}
\label{eq:yII}
\end{equation}

Equations~(7) and (8) provide exact expressions for the experimentally observable responses, explicitly accounting for the finite temporal resolution. Details of the analytic derivation are presented in the Appendix A. Notably, the convolution leads to nonzero signal contributions at negative delay times, a mathematically unavoidable consequence of the Gaussian IRF. 

To illustrate the effect of finite temporal resolution, Fig.~1 compares representative intrinsic and observable responses for the type-I and type-II models. 
For visualization purposes, a constant baseline is added to the normalized signals shown in the figures. 
Representative responses were calculated using $\tau = 1.5$ ps and $\tau_{\mathrm{rec}} = 20$ ps for the type-I model, and $\tau_{\mathrm{fast}} = 0.3$ ps, $\tau_{\mathrm{slow}} = 3$ ps, and $\tau_{\mathrm{rec}} = 20$ ps for the type-II model, with an IRF FWHM of 0.3 ps. The amplitudes were chosen as $A = 0.3$ for the type-I model and $A_{\mathrm{fast}} = 0.2$, $A_{\mathrm{slow}} = 0.1$ for the type-II model. While the intrinsic responses exhibit qualitatively distinct early-time dynamics, convolution with the Gaussian IRF substantially reduces their observable differences. 
\begin{figure}[H]
\centering
\includegraphics[width=\columnwidth]{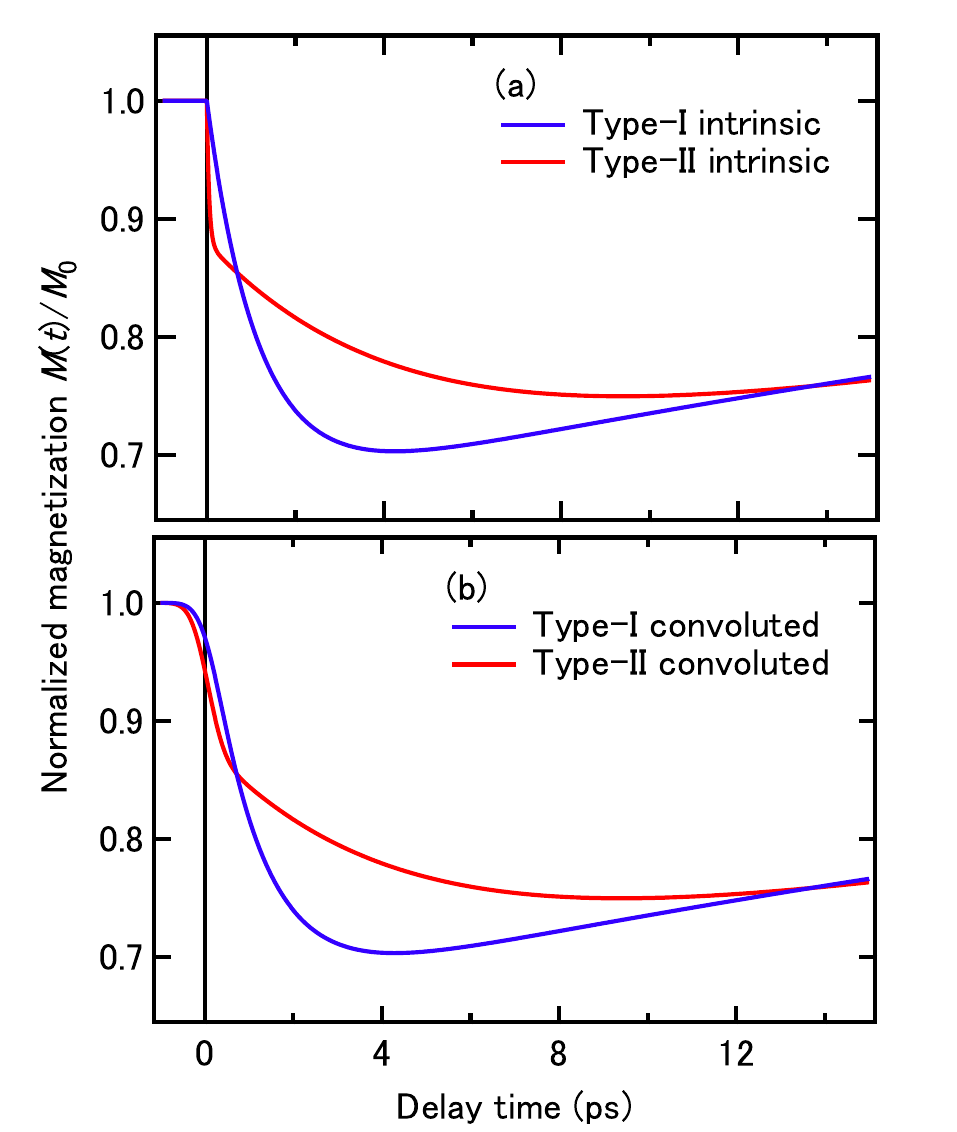}
\caption{(Color online) (a) Intrinsic demagnetization responses for the type-I and type-II models defined in Eqs. (1) and (2). (b) Corresponding observable signals obtained after convolution with a Gaussian instrumental response function (IRF), as described by Eqs. (7) and (8). Finite temporal resolution substantially reduces the observable differences between the two models. Representative parameters are given in the text.}
\label{fig:fig1}
\end{figure}

\subsection{Synthetic data generation}
To assess the distinguishability of type-I and type-II
demagnetization dynamics under realistic experimental
conditions, we generate synthetic pump-probe datasets
based on the IRF-convolved response functions defined
in Sec.~II. The time delay $t$ is sampled on a uniform grid,
\begin{equation}
t_i = t_{\min} + i \Delta t,
\end{equation}
where \(t_{\min}\) defines the earliest negative delay, \(\Delta t\) is the temporal step size, and \(i\) is an integer index. Unless otherwise stated, the synthetic datasets presented below are generated from the type-II model, which serves as the ground truth in the classification
problem.

To emulate experimental uncertainties, statistically independent Gaussian noise is added to the noiseless signal, 
\begin{equation}
d_i = y(t_i) + \varepsilon_i,
\end{equation}
where
\begin{equation}
\varepsilon_i \sim \mathcal{N}(0,\sigma_n^2).
\end{equation}
The noise amplitude \(\sigma_n\) is chosen to be
representative of a typical ultrafast pump-probe
measurements. These synthetic datasets provide a
controlled benchmark for evaluating the model
distinguishability under finite temporal resolution and
noise.

\subsection{BIC-based model comparison}
To quantitatively compare the type-I and type-II
models, we employ the Bayesian information criterion
(BIC)~\cite{Schwarz1978}, which provides a large-sample
approximation to Bayesian model selection~\cite{MacKay2003}.

Given time-resolved data
\begin{equation}
D = \{ d_i \equiv d(t_i) \}_{i=1}^{N},
\end{equation}
and a model prediction \(y(t_i;\theta,M)\), we assume
statistically independent Gaussian noise with variance
\(\sigma_n^2\). The likelihood function is therefore

\begin{equation}
p(D|\theta, M)
=\prod_{i=1}^{N}\frac{1}{\sqrt{2\pi}\sigma_n}
\exp\left[-\frac{(d_i-y(t_i;\theta,M))^2}
{2\sigma_n^2}
\right].
\end{equation}

Conventional least-squares fitting corresponds to
maximizing this likelihood with respect to the model
parameters \(\theta\). However, least-squares fitting
alone does not provide a quantitative basis for
comparing models with different structures.

The BIC is defined as
\begin{equation}
\mathrm{BIC}=k\ln N-2\ln\hat{L},
\end{equation}
where \(k\) is the number of fitting parameters,
\(N\) is the number of data points, and \(\hat{L}\)
is the maximum likelihood. For the present phenomenological models, $k=3$ ($A$, $\tau$, and $\tau_{\mathrm{rec}}$) for the type-I model and $k = 5$ ($A_{\mathrm{fast}}$, $A_{\mathrm{slow}}$, $\tau_{\mathrm{fast}}$, $\tau_{\mathrm{slow}}$, and $\tau_{\mathrm{rec}}$) for the type-II model.

To compare the type-I and type-II models, we evaluate an approximate Bayes factor derived from the BIC difference,

\begin{equation}
K_{\mathrm{BIC}}^{\mathrm{II/I}}
=\exp\left[-\frac{\mathrm{BIC}_{\mathrm{II}}-
\mathrm{BIC}_{\mathrm{I}}}{2}\right].
\end{equation}
Positive values of $\log_{10} K_{\mathrm{BIC}}^{\mathrm{II/I}}$ favor the type-II model, whereas negative values favor the type-I model. Values near zero indicate statistically inconclusive regimes.

\section{Results}
We now present the results of Bayesian model comparison applied to the synthetic datasets introduced in Sec.~II C. The focus is placed on how the distinguishability between type-I and type-II dynamics depends on the temporal resolution and noise level, which are key experimental limitations in ultrafast pump-probe measurements.

\subsection{Least-squares fitting of representative synthetic datasets}
We first consider a representative synthetic dataset generated from the type-II model using 
$\tau_{\mathrm{fast}} = 0.3$ ps,
$\tau_{\mathrm{slow}} = 3$ ps, 
$\tau_{\mathrm{rec}} = 20$ ps, 
$A_{\mathrm{fast}}= 0.2$, 
and $A_{\mathrm{slow}} = 0.1$,
with an IRF FWHM of 0.3 ps and a noise amplitude $\sigma_n = 0.01$.

Figure 2(a) compares the synthetic data with the least-squares fits obtained using the type-I and type-II models.

Despite the intrinsically two-step nature of the underlying dynamics, both models reproduce the observable signal reasonably well within the noise level. In particular, the finite instrumental response smooths the early-time curvature, making the two phenomenological descriptions visually similar over much of the measured delay range.

However, the residual analysis shown in Fig.~2(b) reveals systematic deviations for the type-I model, especially in the early-time region. In contrast, the residuals associated with the type-II model remain more randomly distributed around zero. These results indicate that conventional least-squares fitting alone may be insufficient to discriminate between the two models under realistic experimental conditions unambiguously.

\begin{figure}[H]
\centering
\includegraphics[width=\columnwidth]{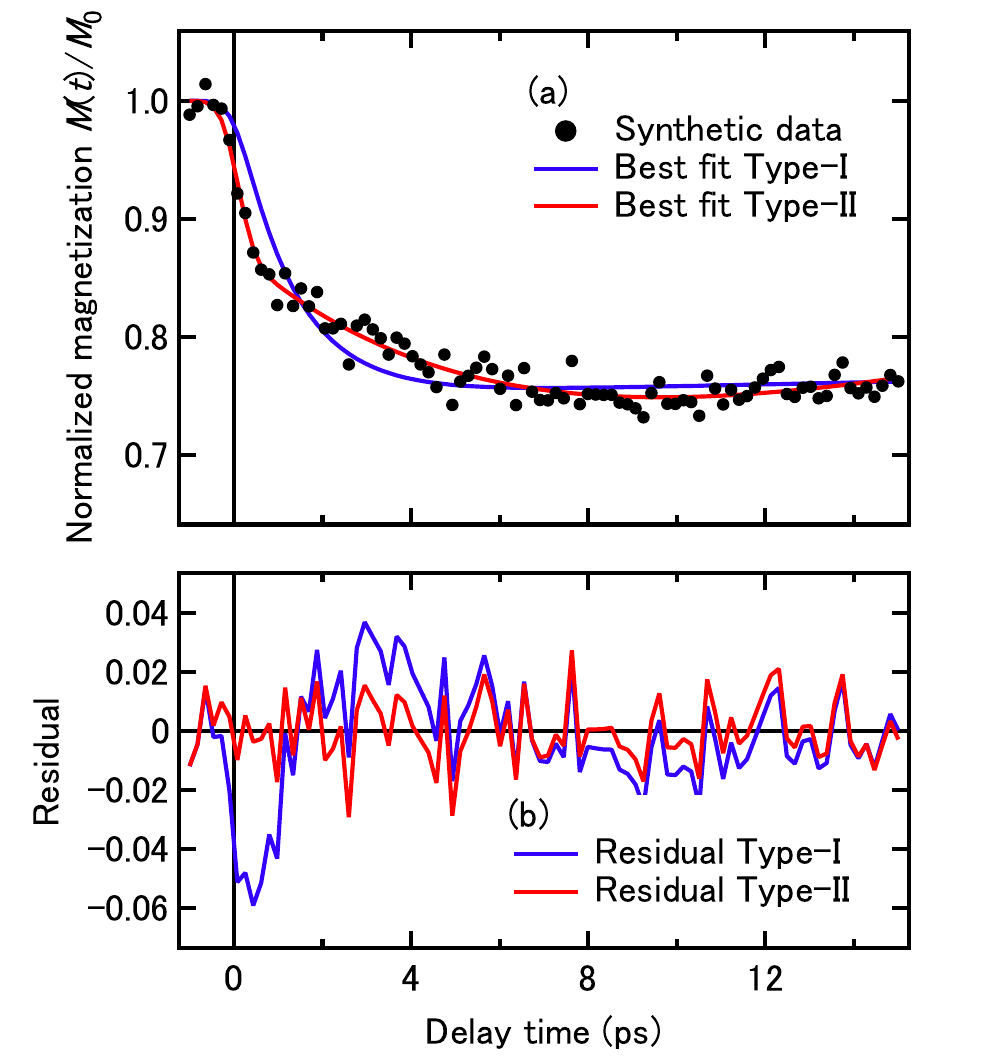}
\caption{(Color online) (a) Synthetic pump–probe data generated from the type-II model (symbols), together with best-fit curves obtained using the type-I and type-II models based on least-squares fitting. (b) Residuals corresponding to the fits shown in (a). The type-I model exhibits systematic deviations at early delay times, whereas the type-II residuals remain more uniformly distributed around zero.}
\label{fig:fig2}
\end{figure}

\subsection{Dependence on temporal resolution and noise}
To systematically investigate the conditions under which the two models become distinguishable, we evaluated the BIC-based approximate Bayes factor over a wide range of temporal resolutions and noise amplitudes.


Figure 3 presents the resulting model preference map as a function of the IRF full width at half maximum (FWHM) and the noise amplitude $\sigma_n$. Positive values of $\log_{10}{K^{\mathrm{BIC}}_\mathrm{{II/I}}}$ indicate a statistical preference for the type-II model, whereas negative values favor the type-I model. Values near zero correspond to experimentally inconclusive regimes in which the available data do not provide meaningful discrimination between the two models.

\begin{figure}[H]
\centering
\includegraphics[width=\columnwidth]{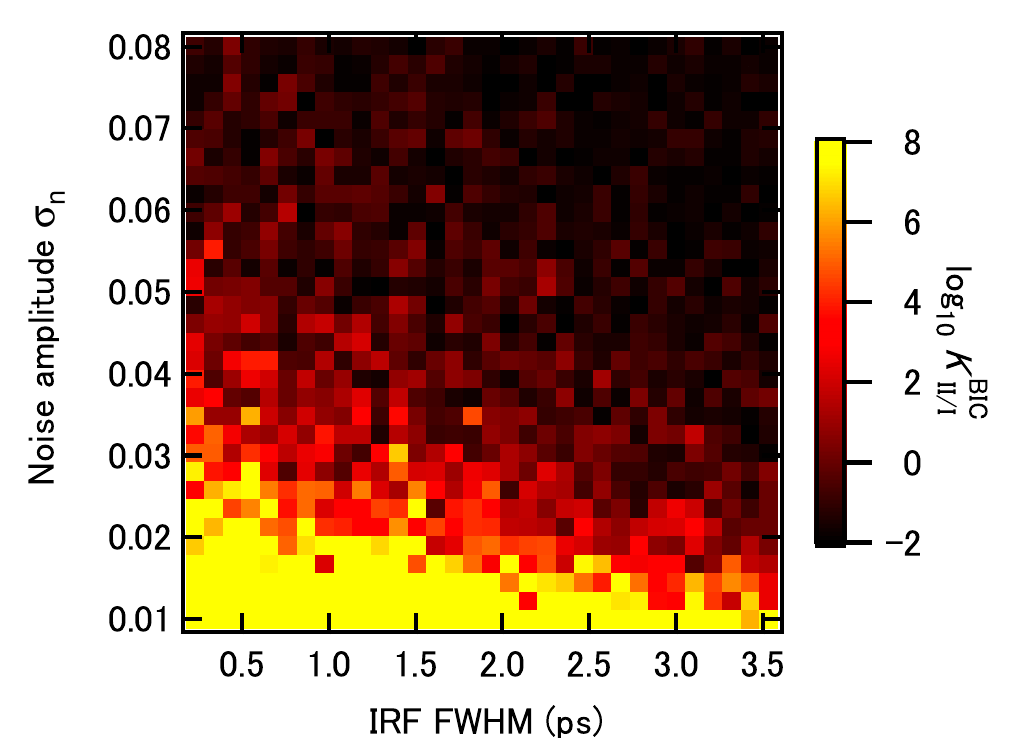}
\caption{(Color online) BIC-based model preference map as a
function of the IRF full width at half maximum (FWHM)
and the noise amplitude $\sigma_n$. 
}
\label{fig:fig3}
\end{figure}

For sufficiently small IRF widths and low noise levels, the type-II model is strongly favored because the intrinsic two-step curvature remains partially resolvable. However, increasing temporal broadening and noise progressively reduce the distinguishability between the two models, eventually leading to statistically inconclusive regimes.

Importantly, this crossover occurs despite fixed intrinsic type-II dynamics, demonstrating that finite experimental resolution alone can fundamentally limit the interpretability of phenomenological classifications in ultrafast demagnetization measurements.

\subsection{Application to experimental data}
To demonstrate the applicability of the proposed framework beyond synthetic benchmark datasets, we further analyzed representative ultrafast demagnetization data for NiCo$_2$O$_4$ thin films reported in Refs.~\cite{10,11,12}. NiCo$_2$O$_4$ is a ferrimagnetic spinel oxide that has recently attracted attention as a candidate material exhibiting complex ultrafast demagnetization dynamics. Previous time-resolved magneto-optical Kerr measurements have suggested a two-step demagnetization process \cite{13}, making this material a suitable test case for the present model-selection framework.

Figure 4(a) compares the best-fit curves obtained from the type-I and type-II models. While both models reproduce the overall temporal evolution reasonably well, systematic differences become visible in the early-time and long-delay regions.

\begin{figure}[H]
\centering
\includegraphics[width=\columnwidth]{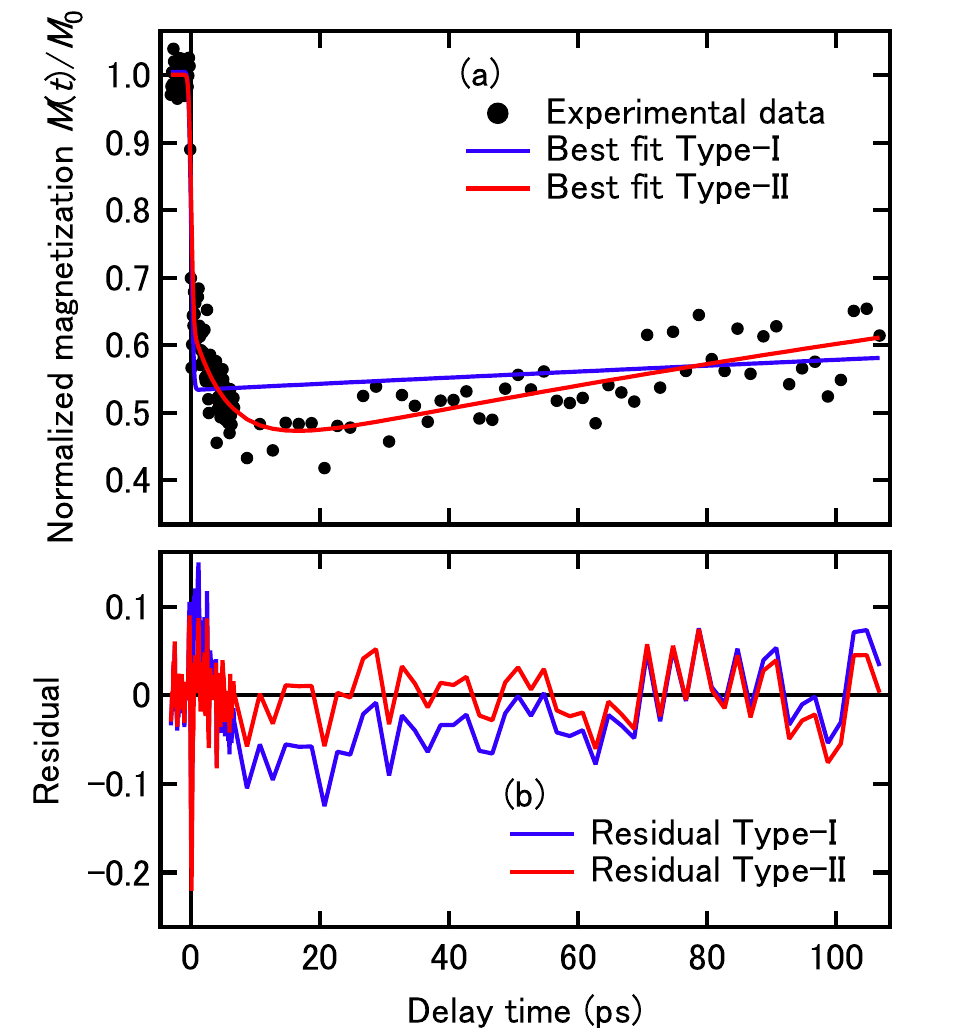}
\caption{(Color online) (a) Representative experimental pump–probe data (symbols) analyzed using the phenomenological type-I and type-II models. (b) Residuals corresponding to the fits shown in (a). The type-I model exhibits more structured residuals over an extended delay range, whereas the type-II residuals remain more uniformly distributed around zero.}
\label{fig:fig4}
\end{figure}

The corresponding residuals are shown in Fig.~4(b). The type-I model exhibits structured residuals extending over a broad delay range, whereas the residuals associated with the type-II model remain more uniformly distributed around zero. This behavior suggests that the additional dynamical component included in the type-II model captures features not accounted for within the simpler single-step description.
The BIC analysis yields $\log_{10}K_{\rm BIC}^{\rm II/I}=82.4$, indicating overwhelming statistical preference for the type-II model. The extracted slow demagnetization time constant ($\tau_{\rm slow}=5.74$ ps) agrees well with the value of approximately 5-6 ps reported in the previous experimental study of NiCo$_2$O$_4$ \cite{13}. This agreement supports the physical consistency of the present fitting framework. To assess the influence of the fitting window, the analysis was repeated using only delay times up to 15 ps. This result indicates that the evidence for type-II dynamics originates primarily from the early-time demagnetization behavior. The resulting model comparison still yielded a strongly positive value of $\log_{10}K_{\rm BIC}^{\rm II/I}=53.7$, indicating that the preference for the type-II model is not solely determined by the long-delay recovery dynamics.

The experimental analysis suggests that Bayesian-inspired model comparison may provide a useful framework for evaluating competing interpretations of ultrafast demagnetization dynamics under realistic experimental conditions. 


\section{Discussion}
Taken together, these results demonstrate that the classification of ultrafast demagnetization dynamics into type-I or type-II categories is inherently conditional on experimental resolution and noise. Even when the intrinsic dynamics are unambiguously two-step, finite temporal resolution and realistic noise levels can render the data statistically consistent with simpler single-step models.

The Bayesian framework provides a principled means of quantifying this compatibility through the Bayes factor, thereby replacing qualitative or heuristic classifications with a probabilistic assessment. In regimes where the Bayes factor approaches unity, e.g. $\log_{10}{K^{\mathrm{BIC}}_\mathrm{{II/I}}}=0$, the data do not support a definitive assignment to either class, and any such classification should be regarded as inconclusive.


\subsection{Conditional nature and experimental implications}
A central outcome of this work is that the distinction between type-I and type-II responses is inherently conditional. Even when the intrinsic dynamics follow a clear two-step (type-II) form, finite temporal resolution and realistic noise levels can render the observable signal statistically compatible with a simpler single-step (type-I) model. In such regimes, the data do not support a definitive classification.

This observation suggests that type-I and type-II labels should not be interpreted as absolute properties of a material system, but rather as hypotheses whose plausibility depends on the quality and scope of the available data. From this perspective, apparent discrepancies in the literature regarding reported demagnetization types may partly reflect differences in experimental resolution and analysis strategies rather than genuine physical inconsistencies.

The present results have direct implications for the interpretation of ultrafast pump-probe experiments. First, they emphasize the need to consider the effective temporal resolution when carefully assigning demagnetization types. Second, they suggest that reporting a single best-fit model without a quantitative assessment of alternative hypotheses may be insufficient, particularly in marginal regimes.

The Bayesian approach outlined here can be straightforwardly extended to experimental datasets and more elaborate noise models. While the present study focuses on synthetic data to isolate methodological effects, the framework is general and can accommodate additional physical constraints or microscopic modeling as needed.

\subsection{Limitations of conventional fitting approaches}
Conventional analyses of ultrafast demagnetization dynamics often rely on least-squares fitting with phenomenological models. While such approaches are effective for parameter estimation within a given model, they do not provide a quantitative basis for comparing models with different structures.

As demonstrated in Fig.~2, type-I and type-II models can yield nearly indistinguishable fits and residuals under realistic conditions, even when the underlying dynamics are known to be two-step. In these cases, model selection based solely on visual inspection or goodness-of-fit metrics is inherently ambiguous and may lead to overinterpretation of the fitted parameters.

\subsection{Advantages and outlook of Bayesian model comparison}
The Bayesian framework adopted in this work addresses these limitations by explicitly treating model selection as a probabilistic inference problem \cite{XX}. By evaluating the model evidence, Bayesian model comparison naturally balances goodness of fit against model complexity, implementing an Occam-type penalty without the need for ad hoc criteria.

Importantly, the Bayes factor provides a continuous measure of model preference, allowing one to identify not only regimes in which one model is strongly favored but also regimes in which the data are genuinely inconclusive. In the latter case, the Bayesian analysis correctly reflects the limited information content of the data, rather than forcing a binary classification.

It is important to stress that the phenomenological models considered in this work are not intended to provide a microscopic description of demagnetization processes. Rather, they serve as test cases for assessing the limits of model distinguishability in time-resolved measurements. The proposed framework is therefore complementary to microscopic theories and can be used to evaluate the warranted level of model complexity for a given dataset.

More broadly, the approach developed here illustrates how Bayesian inference can formalize long-standing qualitative distinctions in ultrafast dynamics. Replacing heuristic classifications with probabilistic statements provides a transparent and reproducible basis for comparing competing dynamical hypotheses.

\section{Conclusion}
In this work, we developed a Bayesian framework for distinguishing type-I and type-II ultrafast demagnetization dynamics from time-resolved data. Using Gaussian-convolved phenomenological models and synthetic pump-probe datasets, we demonstrated that finite temporal resolution and realistic noise levels can substantially reduce the distinguishability between the two classes, even when their intrinsic dynamics are qualitatively different.

The present results highlight the importance of statistical model comparison in interpreting ultrafast demagnetization experiments. Bayesian model comparison provides a quantitative criterion for assessing whether available data genuinely support one dynamical model over another or remain statistically inconclusive. The framework developed here can be extended to experimental datasets and more microscopically motivated models in future studies.

\section*{Acknowledgments}
This work was supported by the JSPS KAKENHI under Grant Nos. JP23K25805 and JP25H01251, and the MEXT Quantum Leap Flagship Program (MEXT Q-LEAP) under Grant No. JPMXS0118068681. The source code used for synthetic data generation, least-squares fitting, and BIC analysis is available at https://github.com/hwadati/Type1or2. 

\section*{Appendix A: Derivation of the Gaussian-convolved response functions}

Here we derive Eqs.~(7) and (8) in the main text. We first consider
the convolution of a single exponential decay with a Gaussian
instrumental response function,
\begin{equation}
G_{\sigma}(t)
=
\frac{1}{\sqrt{2\pi}\sigma}
\exp\left(
-\frac{t^2}{2\sigma^2}
\right).
\end{equation}
For an exponential response that starts at \(t=0\), the basic
convolution integral is
\begin{equation}
\Phi(t;\lambda,\sigma)
=
\int_{0}^{\infty}
e^{-u/\lambda}
G_{\sigma}(t-u)\,du .
\end{equation}
Substituting the Gaussian form gives
\begin{equation}
\Phi(t;\lambda,\sigma)
=
\frac{1}{\sqrt{2\pi}\sigma}
\int_{0}^{\infty}
\exp\left[
-\frac{u}{\lambda}
-\frac{(t-u)^2}{2\sigma^2}
\right]du .
\end{equation}
Completing the square in the exponent yields
\begin{equation}
-\frac{u}{\lambda}
-\frac{(t-u)^2}{2\sigma^2}
=
-\frac{1}{2\sigma^2}
\left[
u-\left(t-\frac{\sigma^2}{\lambda}\right)
\right]^2
+
\frac{\sigma^2}{2\lambda^2}
-
\frac{t}{\lambda}.
\end{equation}
Therefore,
\begin{equation}
\begin{aligned}
\Phi(t;\lambda,\sigma)
&=
\frac{1}{\sqrt{2\pi}\sigma}
\exp\!\left(
\frac{\sigma^2}{2\lambda^2}
-\frac{t}{\lambda}
\right)
\\
&\times
\int_0^\infty
\exp\!\left[
-\frac{
\left\{
u-\left(t-\frac{\sigma^2}{\lambda}\right)
\right\}^2
}
{2\sigma^2}
\right]du
\end{aligned}
\end{equation}
Using the change of variable
\begin{equation}
z =
\frac{
u-\left(t-\sigma^2/\lambda\right)
}{\sqrt{2}\sigma},
\end{equation}
we obtain
\begin{equation}
\Phi(t;\lambda,\sigma)
=
\frac{1}{2}
\exp\left(
\frac{\sigma^2}{2\lambda^2}
-
\frac{t}{\lambda}
\right)
\operatorname{erfc}
\left(
\frac{\sigma^2/\lambda-t}{\sqrt{2}\sigma}
\right),
\end{equation}
which corresponds to Eq.~(6) in the main text.

The type-I intrinsic response can be written as
\begin{equation}
M_{\mathrm{I}}(t)
=
1
-
A
\left(
1-e^{-t/\tau}
\right)
e^{-t/\tau_{\mathrm{rec}}}.
\end{equation}
Expanding the product gives
\begin{equation}
M_{\mathrm{I}}(t)
=
1
-
A
\left[
e^{-t/\tau_{\mathrm{rec}}}
-
e^{-t/\tau_{\mathrm{eff}}}
\right],
\end{equation}
where
\begin{equation}
\frac{1}{\tau_{\mathrm{eff}}}
=
\frac{1}{\tau}
+
\frac{1}{\tau_{\mathrm{rec}}},
\qquad
\tau_{\mathrm{eff}}
=
\frac{\tau\tau_{\mathrm{rec}}}
{\tau+\tau_{\mathrm{rec}}}.
\end{equation}
Using the linearity of convolution, the Gaussian-convolved
type-I response is therefore
\begin{equation}
y_{\mathrm{I}}(t)
=
1
-
A
\left[
\Phi(t;\tau_{\mathrm{rec}},\sigma)
-
\Phi\left(
t;
\frac{\tau\tau_{\mathrm{rec}}}
{\tau+\tau_{\mathrm{rec}}},
\sigma
\right)
\right].
\end{equation}

Similarly, the type-II intrinsic response is
\begin{equation}
\begin{aligned}
M_{\mathrm{II}}(t)
=
1-
\Big[
&A_{\mathrm{fast}}
\left(1-e^{-t/\tau_{\mathrm{fast}}}\right)
\\
&+
A_{\mathrm{slow}}
\left(1-e^{-t/\tau_{\mathrm{slow}}}\right)
\Big]
e^{-t/\tau_{\mathrm{rec}}}.
\end{aligned}
\end{equation}
Expanding each term and applying the same convolution identity
gives
\begin{align}
y_{\mathrm{II}}(t)
=
1
&-
A_{\mathrm{fast}}
\left[
\Phi(t;\tau_{\mathrm{rec}},\sigma)
-
\Phi\left(
t;
\frac{\tau_{\mathrm{fast}}\tau_{\mathrm{rec}}}
{\tau_{\mathrm{fast}}+\tau_{\mathrm{rec}}},
\sigma
\right)
\right]
\nonumber\\
&-
A_{\mathrm{slow}}
\left[
\Phi(t;\tau_{\mathrm{rec}},\sigma)
-
\Phi\left(
t;
\frac{\tau_{\mathrm{slow}}\tau_{\mathrm{rec}}}
{\tau_{\mathrm{slow}}+\tau_{\mathrm{rec}}},
\sigma
\right)
\right].
\end{align}
These expressions give the analytic Gaussian-convolved response functions used as forward models in the main text.

\section*{Appendix B: Additional fitting results}

\begin{table}[H]
\caption{Best-fit parameters obtained for the full-range experimental data shown in Fig.~4.}
\label{tab:fitparam}
\centering
\begin{tabular}{lcc}
\hline
Parameter & Type-I & Type-II \\
\hline
$\tau$ (ps) & 0.05 & -- \\
$\tau_{\rm rec}$ (ps) & 1000 & 278 \\
$\tau_{\rm fast}$ (ps) & -- & 0.05 \\
$\tau_{\rm slow}$ (ps) & -- & 5.74 \\
\hline
BIC & 936.0 & 556.3 \\
\hline
\multicolumn{3}{c}{$\log_{10}K_{\rm BIC}^{\rm II/I}=82.4$} \\
\hline
\end{tabular}
\end{table}


\begin{thebibliography}{99}
\bibitem{1}
E. Beaurepaire, J.-C. Merle, A. Daunois, and J.-Y. Bigot, “Ultrafast Spin Dynamics in Ferromagnetic Nickel,” Phys. Rev. Lett. {\bf 76}, 4250 (1996).

\bibitem{2}
A. Kirilyuk, A. V. Kimel, and T. Rasing, “Ultrafast optical manipulation of magnetic order,” Rev. Mod. Phys. {\bf 82}, 2731 (2010).

\bibitem{3} 
B. Koopmans, G. Malinowski, F. D. Longa, D. Steiauf, M. Fahnle, T. Roth, M. Cinchetti, and M. Aeschlimann, “Explaining the paradoxical diversity of ultrafast laser-induced demagnetization", Nat. Mater. {\bf 9}, 259 (2010).

\bibitem{4} 
A. J. Schellekens and B. Koopmans,
“Comparing Ultrafast Demagnetization Rates Between Competing Models,” Phys. Rev. Lett. {\bf 110}, 217204 (2013).

\bibitem{5}
M. Battiato, K. Carva, and P. M. Oppeneer, “Superdiffusive Spin Transport as a Mechanism of Ultrafast Demagnetization,” Phys. Rev. Lett. {\bf 105}, 027203 (2010).

\bibitem{mentik}
J. H. Mentink, J. Hellsvik, D. V. Afanasiev, B. A. Ivanov, A. Kirilyuk, A. V. Kimel, O. Eriksson, M. I. Katsnelson, and Th. Rasing, “Ultrafast Spin Dynamics in Multisublattice Magnets," Phys. Rev. Lett. 108, 057202 (2012).

\bibitem{Schwarz1978}
G. Schwarz, “Estimating the Dimension of a Model,” Ann. Statist. {\bf 6}, 461 (1978).

\bibitem{MacKay2003}
D. J. C. MacKay, Information Theory, Inference, and Learning Algorithms (Cambridge University Press, Cambridge, England, 2003).

\bibitem{10}
R. Takahashi, Y. Tani, H. Abe, M. Yamasaki, I. Suzuki, D. Kan, Y. Shimakawa, and H. Wadati, “Ultrafast demagnetization in NiCo$_2$O$_4$ thin films probed by time-resolved microscopy,” Appl. Phys. Lett. {\bf 119}, 102404 (2021). 

\bibitem{11}
R. Takahashi, T. Ohkochi, D. Kan, Y. Shimakawa, and
H. Wadati, “Optically induced magnetization switching in NiCo$_2$O$_4$ thin films using ultrafast lasers", ACS Appl. Electron. Mater. {\bf 5}, 748 (2023).

\bibitem{12}
R. Takahashi, Y. Le Guen, S. Nakata, J. Igarashi, J. Hohlfeld, G. Malinowski, L. Xie, D. Kan, Y. Shimakawa, S. Mangin, and H. Wadati, “All-optical helicity-dependent switching in NiCo$_2$O$_4$ thin films", Appl. Phys. Lett. {\bf 126}, 212405 (2025).

\bibitem{13}
R. Takahashi, K. Yamada, H. Singh, K. Watanabe, J. Igarashi, J. Hohlfeld, J. Gorchon, G. Malinowski, D. Kan, Y. Shimakawa, T. Ishibashi, S. Mangin, and H. Wadati, “Type-II-like ultrafast demagnetization behavior in NiCo$_2$O$_4$ thin films", arXiv:2604.17916v1. 

\bibitem{XX}
C. M. Bishop, Pattern Recognition and Machine Learning, Springer (2006).

\bibitem{YY}
H. Shinotsuka, K. Nagata, H. Yoshikawa, Y. Mototake, H. Shouno, and M. Okada, “Development of spectral decomposition based on Bayesian information criterion with estimation of confidence interval", Sci. Technol. Adv. Mater. {\bf 21}, 402 (2020). 

\bibitem{ZZ}
A. R. Liddle, "Information criteria for astrophysical model selection", Mon. Not. R. Astron. Soc. {\bf 377}, L74 (2007).

\bibitem{AA}
J. K. Webb, C.-C. Lee, R. F. Carswell and D. Milaković, "Getting the model right: an information criterion for spectroscopy", Mon. Not. Roy. Astron. Soc. {\bf 501}, 2268 (2021). 


\end{thebibliography}
\end{document}